# Data-Driven Spectrum Access Technologies for 6G

Foad Sohrabi, Harish Viswanathan, Sivarama Venkatesan, Jinfeng Du, Aliye Özge Kaya, Navin Hathiramani, and Pavan Koteshwar Srinath

*Abstract*— **The surge in global mobile traffic, driven by 5G adoption and emerging applications, necessitates new spectrum allocations for 6G systems. With most major swaths of spectrum below 15 GHz already allocated to other services, co-existence between cellular and incumbent services such as satellite or radar systems is critical to make such spectrum accessible for 6G. At the same time, the advent of artificial intelligence offers a new opportunity for designing more efficient spectrum sharing techniques. This paper presents four novel data-driven spectrum access techniques to enable efficient use of spectrum for the 6G era: 1) co-existence technique for efficient protection of fixed incumbent receivers within an established protection area; 2) data-driven approach for protecting satellite receivers via control of radiated power above the horizon; 3) spectrum sensing-based fast-time-scale physical resource block blanking; and 4) base station-based spectrum occupancy detection to enable fast-time-scale spectrum sharing with satellite and non-terrestrial network incumbents. These techniques leverage artificial intelligence and extreme multiple-input multiple-output capabilities to enable efficient spectrum co-existence/sharing while maintaining cellular network performance.**

## I. Introduction

GLOBAL mobile traffic continues its exponential growth, fueled by accelerated 5G adoption, enhanced streaming video quality, expanding Augmented Reality and Virtual Reality ecosystems, cloud gaming, and AI integration. Rapid growth in overall mobile data volume and the evolving requirements of new generations of technologies, devices, applications, and services has been the main drivers for more spectrum in mobile systems. At the World Radiocommunication Conference in 2023 (WRC-23), the International Telecommunication Union agreed [1] to study the 7.125–8.4 GHz band for potential International Mobile Telecommunications identification as a prime 6G band, with decisions planned for WRC-27. Additionally, the upper 6 GHz band (6.425–7.125 GHz) has been identified for 5G Advanced and 6G with Above-The-Horizon (ATH) power limits for satellite coexistence.

These upper mid-bands offer substantial advantages for 6G deployments: wider-bandwidth carriers, deployment on existing 5G site grids using similar antenna array dimensions, and higher capacity through increased antenna elements and transceiver chains. However, in many countries, large portions of these bands are occupied by incumbent terrestrial and satellite services. The prospect of relocating incumbents to free up the spectrum for 6G may be possible only for some of the incumbents, making spectrum sharing essential.

While sharing spectrum across 5G and 6G technologies within a single service provider network in the to-be re-farmed 5G bands will be addressed by Multi-RAT Spectrum Sharing (MRSS) technology [2] spectrum sharing with other incumbents has been a topic of intense research and development for decades with numerous sharing technologies and policies proposed over the years [3] –[6] .

Citizens Broadband Radio Service (CBRS) spectrum allocation to mobile systems in the US is arguably the most successful example of spectrum sharing. In CBRS, a three-tier sharing framework is established where the incumbents have the highest priority for spectrum usage, followed by priority-access licensees while obeying transmission power limits. With the established framework, the shared spectrum is accessible to wireless deployments when, for example, navy radars are not in use. The CBRS framework for incumbent protection has been evolving and enhancements were recently adopted [6] to account for traffic load and time division duplex patterns as well as more realistic path loss models in aggregated interference calculation. Advanced spectrum sharing techniques for the 6G band are of intense interest and form the main focus of this paper.

While spectrum sharing has always relied on accurate spectrum usage data gathered from databases and sensors, it has not relied on data-driven optimization of the cellular transmitted signal in time, frequency and spatial dimensions. Artificial Intelligence/Machine Learning (AI/ML) are emerging tools that heavily rely on data for training and structure abstraction, thus often called data-driven techniques. They offer new possibilities for spectrum sharing when combined with extreme Multiple-Input Multiple-Output (MIMO) in 6G, where the transmitted signal can be accurately beamformed in certain directions with narrow beamwidth while creating nulls in other directions [7] allowing for more efficient spatial sharing of spectrum with incumbent receivers.

Spectrum sharing between satellite and terrestrial mobile services has some unique challenges. For example, the high demand for spectrum for different services is often in the same densely populated areas at the same time. The satellite beam-frequency association could also be reconfigured based on satellite traffic distributions [8] -[10] . Fortunately, not all spectrum is used in all neighboring satellite beam spots due to frequency reuse among neighboring beams to mitigate inter-beam interference, which provides opportunities for spectrum sharing even when satellite traffic is heavily loaded.

We present four novel spectrum sharing techniques tailored for the 6G era, designed to comply with potential policy frameworks while maximizing spectral usage and efficiency.

*Foad Sohrabi, Harish Viswanathan, Sivarama Venkatesan, Jinfeng Du, and Aliye Özge Kaya are with Nokia Bell Labs, Murray Hill, New Jersey, USA. Navin Hathiramani is with Nokia Standards, Dallas, Texas, USA. Pavan Koteshwar Srinath is with Nokia Bell Labs, Messy, France.*

First, efficient protection of fixed incumbent receivers with known protection areas is achieved through extreme MIMO beamforming enabled by a data-driven optimization technique. Second, a closely related data-driven approach is developed to protect signal reception at satellites from cellular interference by limiting ATH radiated power while maximizing terrestrial cellular network spectral efficiency. Third, Physical Resource Block (PRB) blanking of cellular signals is improved to protect mobile satellite receivers using a combination of satellite trajectory data and spectrum sensing at Base Station (BS) sites. Fourth, BS-based spectrum occupancy detection is proposed as a new paradigm to enhance the efficiency of fast-time-scale spectrum sharing with satellite incumbents.

## II. Spectrum Landscape and Protection Requirements

The spectrum landscape in the US for the 6G band identified for decision-making at WRC-27 is included in Fig. 1, generated based on data from [5] . Different parts of the band are occupied by different types of fixed and mobile satellite services, as well as terrestrial fixed services, assigned with either primary (green) or secondary (orange) user status. To maximize spectrum for future 6G systems, sharing spectrum with incumbents will be needed, where transmissions from BSs and user devices shall not cause harmful interference to incumbents and the cellular system receivers should be resilient to any interference generated by the incumbent systems.

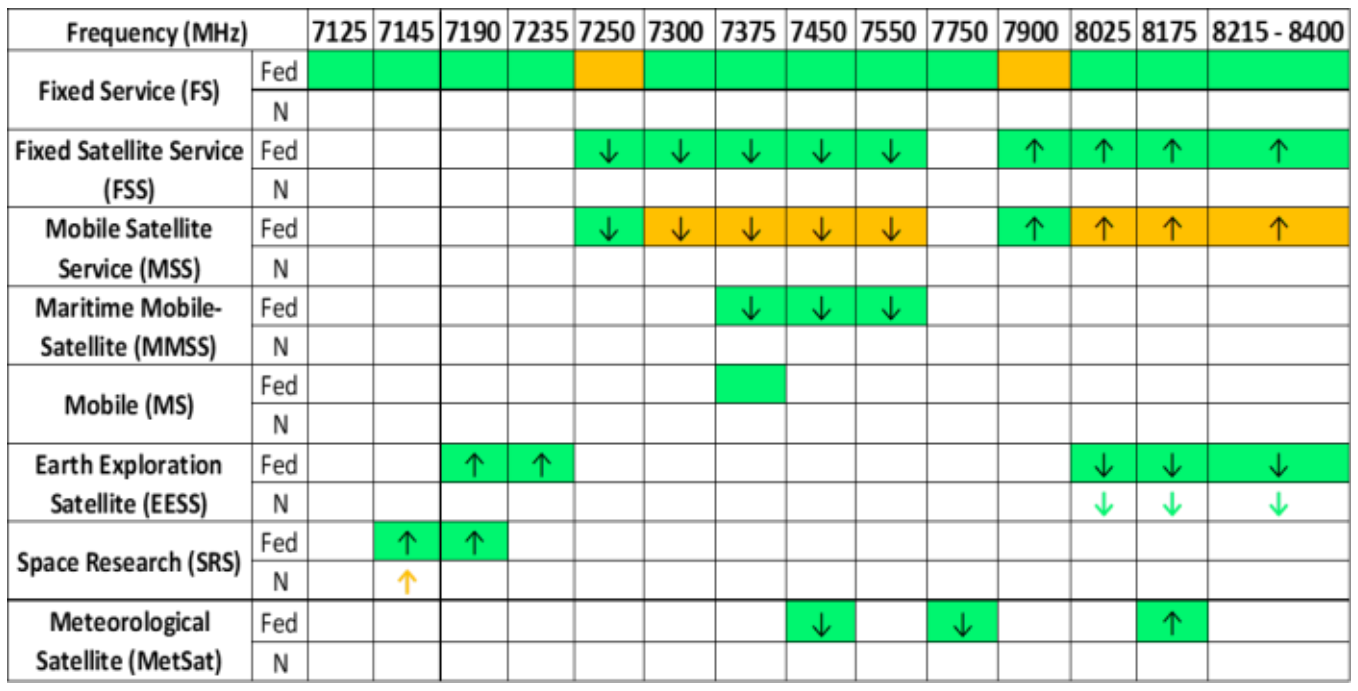

*Figure 1. Spectrum landscape in the US for 7.125–8.4 GHz, generated based on data from [5] , where Earth-to-space (↑) and space-to-Earth (↓) links of satellite services have primary (green) or secondary (orange) allocation for federal (Fed) and non-federal (N) use. Arrows without background color indicate limitation in either frequency range or geographical areas.*

Spectrum sharing for at least three types of coexistence scenarios could be required for the newly identified 6G spectrum to protect the incumbent services, which correspond to three different types of receivers:

- **Stationary terrestrial receivers** whose locations are well known - typically Earth stations of fixed satellite services or receiver part of fixed terrestrial services. Traditional beam muting or power backoff could lead to substantial performance degradation.
- **Mobile terrestrial receivers** whose exact locations are unknown but the transmitting satellite's trajectory or location is known, and the presence of satellite transmission signals could be detected. Fine-granularity fast dynamic spectrum sharing solutions are needed for effective spectrum use.
- **Satellite receivers** that receive signals from Earth stations or other mobile transmitters, typically moving relative to the ground along certain trajectories, such as Low-Earth Orbit (LEO) satellites, or are stationary relative to Earth, such as geostationary satellites like the Wideband Global Satcom (WGS) system [9] . Meeting the ATH radiation limit is a challenge for conventional beamforming in cellular systems.

## III. Beamforming Optimization for Sharing with Incumbents Inside Known Protection Areas

This section focuses on scenarios where the Mobile Network (MN) has access to information about a protection area for incumbents such as satellite ground stations or radars and aims to protect them from harmful interference caused by the MN transmissions. By allowing the BS to obtain samples or models of the Channel State Information (CSI) within the protection area, it can reduce transmit power to ensure that the received power at the incumbent(s) in the protection area remains within acceptable levels. This, however, can lead to significant MN coverage and data rates degradation, particularly in areas with high incumbent protection requirements.

One viable solution to address this issue is to leverage the spatial domain. In the emerging extreme MIMO systems where BSs employ large antenna arrays, beamforming techniques, alongside power control, can play a crucial role in shaping the signal transmission and mitigating interference. However, conventional beamforming methods for sharing among terrestrial systems are targeted to reduce intra-system interference and do not specifically address the challenge of protecting incumbents while maximizing utility in shared spectrum scenarios. This motivates the development of novel beamforming techniques that can effectively control the level of interference experienced by incumbents in a designated protection area, thus enabling the coexistence of MNs and incumbent services without significantly compromising the performance of either.

### *A. Data-Driven Precoding*

To illustrate the concept, consider a single-cell scenario in which a BS with a fully digital, large-antenna array serves a single MN user while controlling its interference toward a protection area for other incumbents. Suppose that the BS has access to the CSI of the served user and to a dataset of channel samples between the BS and points inside the protection area. These samples may be obtained from channel models leveraging terrain and clutter maps, from historical in-band measurements collected when the protection was not active and users in the area were being served, or from real-time measurements on other non-shared frequency resources, with any residual inaccuracy handled using an appropriate power margin. The problem of interest is designing a precoder that maximizes a KPI of the MN (e.g., the user rate) while keeping the interference power at every channel sample inside the protection area below a prescribed threshold. The interference threshold can be set by a spectrum sharing coordinator or controller to allow distributed per-BS interference control while keeping the aggregate interference at a reasonable level, accounting for any margin needed to mitigate inaccuracy in the protection-area channel samples. The spectrum sharing coordinator/controller could be the same entity that provides the MN the needed information about the protection area.

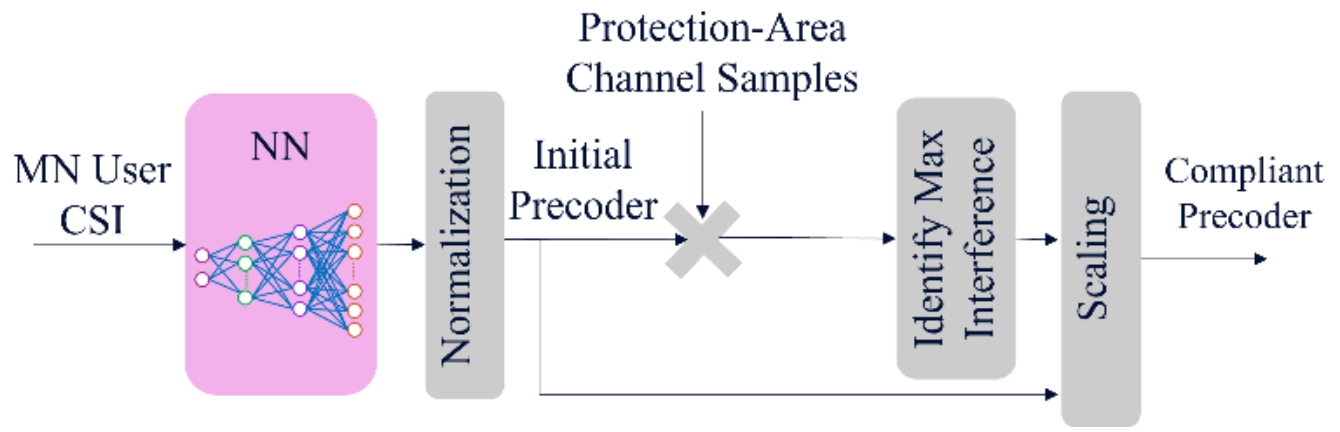

*Figure 2. Data-driven approach for designing codebook-free beamforming for spectrum sharing.*

We consider codebook-free precoding under a per-BS total power constraint. The neural network architecture in Fig. 2 takes both the served-user CSI and the protection-area channel samples as input, and produces a precoder that optimizes the MN utility while satisfying the interference constraint at every sample. In the first stage, a fully-connected neural network (NN) followed by a power normalization layer generates an initial precoder from the served-user CSI. In the second stage, the precoder power is scaled down to ensure that the interference at every sample of the protection-area dataset stays within the prescribed threshold. The training loss is set as the negative of the chosen MN utility (e.g., the user rate). In our implementation, a small fully-connected network with a single hidden layer (e.g., 1024 units for the 128-antenna scenario, tanh activation) is trained end-to-end with Adam, using a fixed set of protection-area channel samples and a separate set of served-user channel realizations split into disjoint training and test subsets.

It should be mentioned that, while the idea of data-driven precoding for spectrum sharing is presented only for codebook-free precoding, this idea can simply be applied for designing a codebook in codebook-based precoding as well. The same data-driven principle also extends to multi-user MIMO transmission. For instance, the free parameter in the generalized-inverse representation of a zero-forcing precoder can be optimized using the protection-area channel sample set to maximize the served-user rates subject to the interference constraint. Alternatively, the dominant left singular vectors of the protection-area sample set can be treated as virtual users and appended to the served-user channel matrix in a classical zero-forcing precoder, with multi-antenna users naturally handled by treating each data stream as a virtual user [11] .

### *B. Numerical Results*

To evaluate the performance of the proposed method, we consider a 128-antenna BS in a UMa scenario at 7 GHz and the following two baselines:

- **Maximum Ratio Transmission** (MRT) + Power Scaling: MRT is known to be optimal codebook-free beamforming for single layer transmission under the sum power constraint. To further make this design compliant with the interference constraint, the power is scaled down.
- **Discrete Fourier Transform** (DFT) Codebook + Power Scaling: The DFT codebook is a widely used codebook for MIMO beamforming. To further make this codebook compliant with the interference constraint, the power of each codeword is properly scaled down.

Fig. 3 illustrates the performance of different methods. The data-driven methods, both codebook-free and codebook-based, significantly outperform their conventional counterparts, confirming that data-driven beamforming designs tailored for spectrum sharing go beyond simple power reduction.

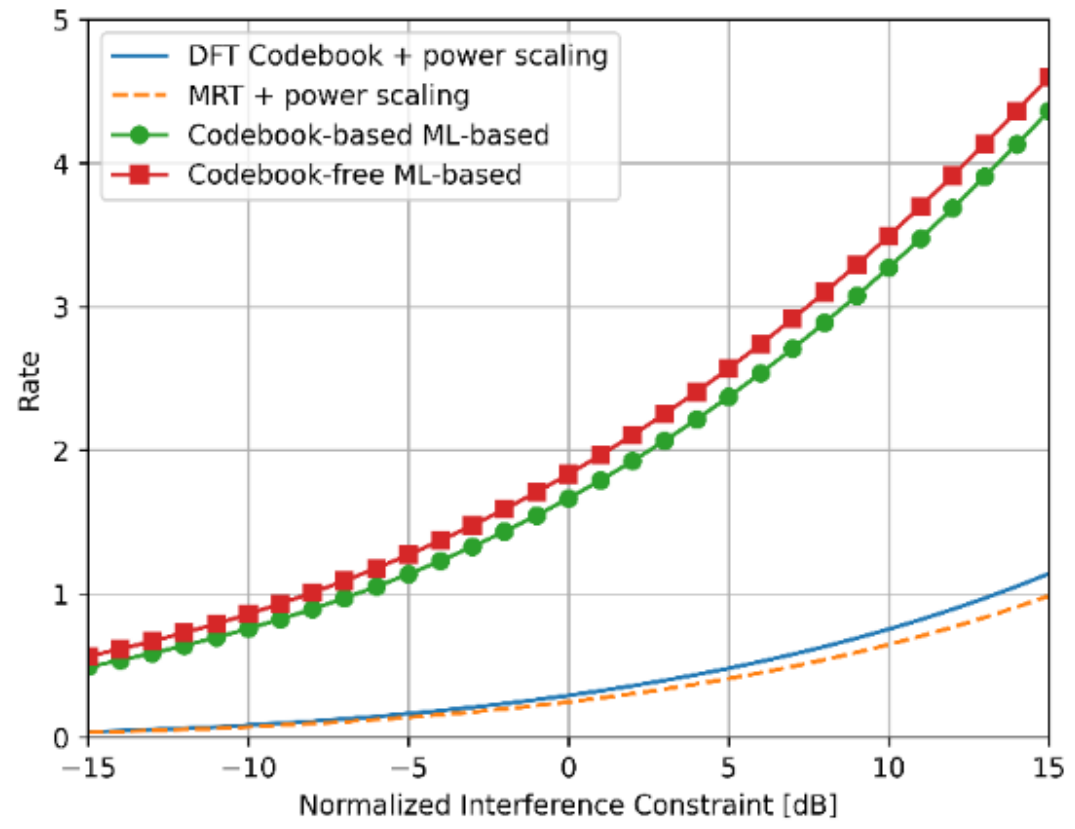

*Figure 3. Average rate versus the normalized interference constraints for different methods.*

## IV. Above-the-Horizon Radiated Power Control for Protecting Satellite Receivers

To control the ATH radiated powers towards satellite incumbents, we assume there is no information about the incumbent location, and the BSs are in unobstructed LOS condition towards satellites. Accordingly, the radiated power constraint should be defined as an average. In the existing systems where the spectrum is mainly allocated to MNs, the BS transmitter configuration and beamforming designs only need to comply with the maximum Effective Isotropic Radiated Power (EIRP) limits. However, new regulations on EIRP limits are being established for new bands that coexist with other networks such as satellite networks. For example, WRC-23 introduced a set of new ATH-EIRP limits [1] as a function of elevation angle for the upper 6-GHz band. These new constraints require the development of novel techniques to design BS configurations and beamforming strategies that meet ATH-EIRP limits with minimum performance degradation.

To implement a massive number of radiators, the extreme MIMO architecture uses a three-stage beamforming process to balance performance with hardware cost, computational complexity, and power consumption. The first two stages, digital precoding and adaptive analog precoding, also known as hybrid beamforming, are based on user CSI. The third stage, called fixed analog precoding, involves fixed phase shifts and attenuations at each radiator, designed during the production phase. Unlike traditional BS designs that use only digital and fixed-analog and stages, the adaptive analog stage enables a smaller number of RF chains to drive a larger radiating array, reducing hardware cost while preserving the massive-MIMO potential. To optimize the system performance while respecting ATH-EIRP constraints, we can design both BS configuration parameters in the production phase as well as the adaptive beamforming strategies.

### *A. Design of BS Configurations*

We use a data-driven approach to tailor the fixed analog beamformer to a given channel distribution, assuming the radiator spacing and gain pattern are fixed. A channel dataset representing users distributed across the topology of interest is collected, for example from a channel generator, ray-tracing tool, or historical measurements. The goal is to find the fixed analog beamformer that maximizes a chosen utility function (e.g., average or cell-edge user rate) while satisfying the ATH-EIRP limit in each elevation bin specified by the regulations. To apply the data-driven approach, we define a loss function inspired by the Lagrangian method, formed as the negative of the utility function plus a weighted sum of the ATH-EIRP constraint violations across all elevation bins.

By selecting a sufficiently large penalty weight on the constraint-violation term, the optimization drives all EIRP violations to zero, achieving ATH-EIRP compliance. With the fixed analog beamformer defined as a trainable parameter and the channel dataset available, stochastic gradient descent (SGD) is used to minimize this loss function and obtain the optimized fixed analog beamformer.

### *B. Numerical Results*

We compare the ML-based solution with solely fixed analog beamforming design to two other methods. The first is the conventional BS design method, in which the fixed analog weights are set by electrical down-tilting: the phase gradients across the array are chosen to steer the main lobe to a fixed elevation angle, when combined with site-specific mechanical down-tilt, pointing toward the most densely populated area in the cell, with no optimization for ATH-EIRP compliance. The second is the conventional filter-based design, which enforces ATH-EIRP limits for almost all possible precoders in a DFT codebook. Fig. 4 shows the cell-average and cell-edge throughput of all designs. Although both ML-based and filter-based designs with 128 TRXs comply with EIRP regulations, ML-based solutions show superior cell-edge performance. Further, the ML approach enables a higher number of TRXs, allowing improved cell average performance by serving more layers with 256 TRXs. This confirms that our proposed framework effectively produces high-quality, EIRP-compliant solutions. The ML-based approach not only meets regulatory requirements but also offers enhanced performance, particularly in challenging network conditions.

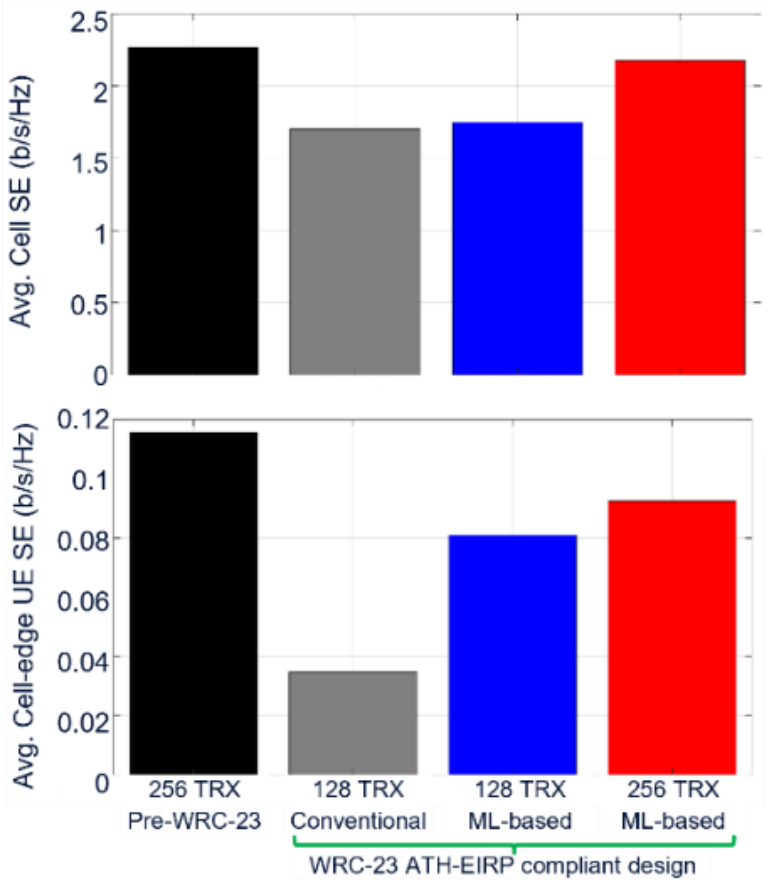

*Figure 4. System performance comparison of different methods.*

## V. Adaptive PRB Blanking for Dynamic Sharing with Mobile Incumbent Receivers

One of the most challenging incumbents to protect is the mobile satellite service devices. Because the incumbent receivers are mobile, it is not possible to protect them by creating static interference-free zones as in the case of fixed receivers. Furthermore, for incumbents whose location is unknown, a possible solution could be to avoid cellular interference in the entire area of the satellite footprint over spectrum where there could potentially be satellite receivers. A potential approach for dynamic spectrum sharing between incumbent satellite services and 6G cellular networks is for cellular BSs to sense and react to the presence of satellite transmissions. Because the satellite signal footprint is large, with the beam spot on the Earth's surface ranging from a radius of tens of kilometers (e.g., LEO satellites [8] ) to hundreds of kilometers (e.g., the WGS system [9] ), there will be a large number of BSs within the footprint area, and the burden of sensing can be shared across the BSs, with some focused on sensing on specific bands, and a selected few targeting wideband sensing, as needed.

Typical satellite services operate a constellation of satellites, which use multiple beams to cover different portions of the Earth's surface. Different frequency carriers are assigned to different beams of a satellite to avoid interference across overlapping beams. Satellite services may partition their allocated spectrum per beam spot. Frequency allocation for satellite coverage in a given geographical area is different from its neighboring satellite cells, and the spectrum occupancy could be dynamically reconfigured [10] [12] depending on traffic demands. Furthermore, only one or two beams typically illuminate each ground location across multiple satellites of the constellation. Hence only a portion of the spectrum allocated to the satellite service is occupied in any given ground location. In addition, if there are no devices being served in the footprint of a beam, the transmission is suppressed in that beam in order to save energy. For these reasons, there is opportunity for BS-sensing based fine-granularity fast dynamic spectrum sharing. For example, when trajectories of satellites are known (e.g., tracked in some database), coverage area-based proactive PRB blanking at cellular BSs could greatly improve the opportunity of spectrum sharing, as shown in [10] , leading to over 60% downlink sum-rate improvement and 110% in uplink for 80% of cases simulated. Another benefit of such geometry-based interfering cell/PRB association is its robustness against potential inaccuracy of LEO trajectories (in between trajectory updates) or BS locations, since the solution is based on geometry rather than CSI.

As soon as the presence of the satellite signal is confirmed by a network of BSs within a beam spot area, all BSs in the footprint of the satellite beam should react to suppress scheduling downlink and uplink transmissions, for example, through adaptive blanking of PRBs corresponding to frequencies occupied by the satellite downlink until the sensing confirms that satellite transmissions have terminated. A combination of proactive and sensing-based reactive solutions could be leveraged to address complex scenarios based on availability of trajectories and frequency allocation of incumbents.

## VI. Base Station-based Spectrum Occupancy Detection for Dynamic Spectrum Sharing

The success of sensing-based dynamic spectrum sharing hinges on reliable spectrum occupancy detection at BSs. An ideal solution should have the following characteristics: (a) no need for *a priori* knowledge of any features of the satellite signal; (b) possibility of decentralized operation at individual base stations; (c) low implementation complexity; (d) ability to sense weak satellite signals reliably; (e) robustness to any fading and shadowing, or the presence of terrestrial interference; and (f) low air interface resource overhead, e.g., for time-frequency blanking to facilitate spectrum occupancy detection.

### A. Energy Detector-based Sensing

A simple *energy detector* would be highly attractive as far as the criteria (a)-(c) above are concerned. Its decision statistic for spectrum occupancy detection is simply the received energy in the bandwidth $B$ of interest over an observation interval of duration $T$, divided by the bandwidth-time product $BT$ (i.e., the received energy per complex dimension in signal space). The sensing BS decides that the satellite signal is present if this statistic exceeds a threshold, and that it is not present otherwise.

The question then is whether (or when) an energy detector would meet the stipulated performance requirements in the criteria (d)-(f) above. More specifically, we are interested in the tradeoff between probabilities of false alarm (i.e., BS declares "present" when it is not) and missed detection (BS decides "not present" when it is), parametrized by the bandwidth-time product and the Signal-to-Noise Ratio (SNR) $\rho$.

Such an assessment was carried out in [13] by treating the satellite signal as deterministic but unknown (apart from the bandwidth that it occupies) and the additive noise as white and Gaussian of known spectral density $N_0$. With a decision threshold of $\lambda$, the false alarm probability $P_{FA}$ was shown to be the probability that a central chi-squared random variable with $2BT$ degrees of freedom exceeds $2BT\lambda/N_0$, and the missed detection probability $P_{MD}$ to be the probability that a non-central chi-squared random variable with $2BT$ degrees of freedom and non-centrality parameter $2BT\rho$ does not exceed $2BT\lambda/N_0$.

For large $BT$, the above probabilities can be well approximated in terms of the Gaussian Q-function using the central limit theorem. Eliminating the decision threshold from these approximations, the SNR required to achieve desired values of $P_{FA}$ and $P_{MD}$ is shown to be inversely proportional to $\sqrt{BT}$ (assuming $BT \gg 1$). This relationship suggests that an energy detector can operate at arbitrarily low SNR, provided the observation bandwidth-time product is large enough.

However, as shown in [14] [15] , this conclusion is crucially dependent on knowing the exact value of the noise power spectral density $N_0$. In practice, there is always some degree of uncertainty in the value of $N_0$, e.g., due to receiver calibration errors or the presence of rapidly changing terrestrial interference (further, the additive noise need not be white or even Gaussian). Following [14] [15] , suppose that the noise spectral density is known only to be within the interval $[N_0/\alpha, \alpha N_0]$, for some $\alpha > 1$. To guarantee a desired false alarm probability $P_{FA}$, the decision threshold must be set to the value required for the *highest* possible noise spectral density, i.e., $\alpha N_0$. To further guarantee a desired missed detection probability $P_{MD}$, the (nominal) SNR $\rho$ must equal or exceed the value required for the *lowest* possible noise spectral density, i.e., $N_0/\alpha$. Eliminating the decision threshold from the central limit approximations for $P_{FA}$ and $P_{MD}$ then yields the relationship

$$\sqrt{BT}(\rho + 1 - \alpha^2) \approx \alpha^2 Q^{-1}(P_{FA}) + (2\rho + 1)Q^{-1}(P_{MD}). \quad (1)$$

Assuming $P_{FA} < 0.5$ and $P_{MD} < 0.5$, this means that *the nominal required SNR* $\rho$ *cannot be smaller than* $\alpha^2 - 1$, *regardless of how large* $BT$ *is.*

### B. Matched-Filter Detector-based Sensing

If exact knowledge of the satellite signal is available at the receiver, a matched-filter detector can be used. It is straightforward to show that, for large $BT$, the SNR $\rho$ required to achieve desired values of $P_{FA}$ and $P_{MD}$ for matched-filter detectors is proportional to $1/(BT)$. Although the assumption of knowing the incumbent signal is unrealistic, its performance could serve as an upper bound on that of more sophisticated spectrum occupancy detection techniques, e.g., those exploiting any known or learned characteristics of the satellite signal using machine learning. With noise uncertainty, the required SNR is given by

$$2BT\rho = \left(\alpha Q^{-1}(P_{FA}) + Q^{-1}(P_{MD})\right)^2. \quad (2)$$

Note that the required SNR $\rho$ remains proportional to $1/(BT)$ even with noise uncertainty -- only the proportionality constant increases.

### C. Performance Evaluation

Fig. 5 shows the required SNR $\rho$ versus the bandwidth-time product $BT$ to achieve false alarm and miss detection rate of 0.01, for different values of $\alpha$ (the dotted lines show the matched filter performance in each case). For reference, the required SNR with no noise uncertainty is also shown (note the slope of $-5$ dB per decade, consistent with $\rho \propto 1/\sqrt{BT}$). The impact of even a small uncertainty in the noise spectral density is quite apparent here, in the form of a floor on the required SNR (below this SNR, it is futile to keep increasing the bandwidth-time product).

The practical implication is that, for spectrum occupancy sensing with an energy detector to be viable, it is important to minimize the uncertainty in the power level of any additive noise or intra-system interference. It would therefore be best to dedicate a small number of the time-frequency resource elements in each frame in every cell exclusively for sensing the possible presence of a satellite signal (this would eliminate intra-system interference while sensing).

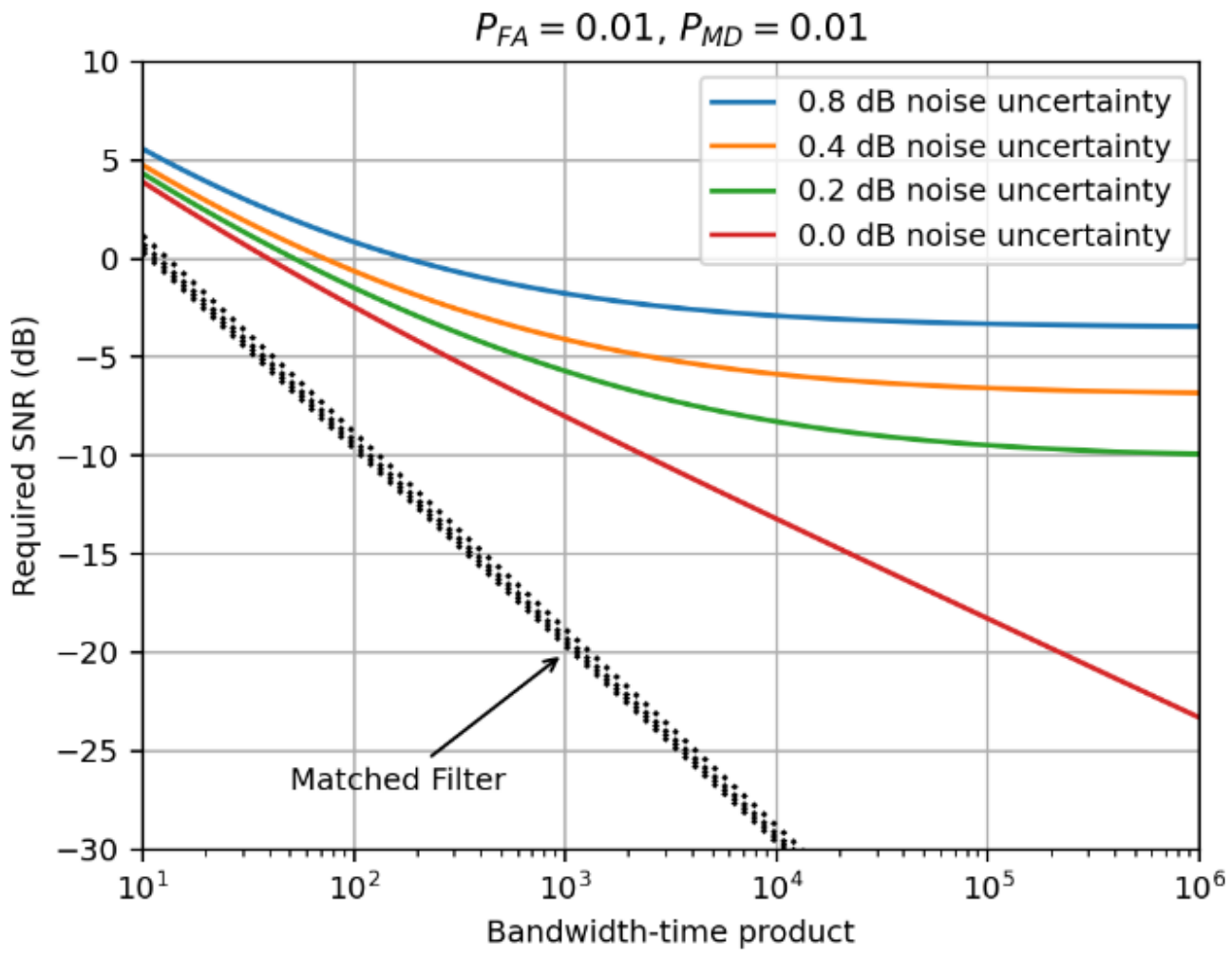

*Figure 5. Required SNR versus bandwidth-time product to achieve 1% false alarm and missed detection probabilities, for different noise uncertainty levels.*

To improve the quality of satellite signal sensing at the BS, uplink signals from UE transmissions could be suppressed, the importance of which has been established here. Note that only a subset of BSs, selected/configured by a spectrum sensing management function, need to skip scheduling uplink transmissions in certain PRBs to facilitate sensing while other BSs may continue to schedule UEs, under certain constraints, to better facilitate spectrum sensing in adjacent spectrum. Network-wide synchronization (to within a few µs) is crucial to enable rapid occupancy sensing with minimal air interface resource overhead.

### *D. BS Traffic Prediction for Sensing Resource Optimization*

The sensing-based dynamic PRB blanking assumes that each network could sense the frequency occupancy within certain areas (as defined by the satellite beam spot). The spectral efficiency of this sharing mechanism is dependent on efficiently enabling resources for incumbent detection in a fast and reliable manner. One method to achieve this could be predicting future PRB usage and employing lower PRB usage periods for sensing.

To have sufficient time to identify candidate BSs and optimize/distribute the spectrum sensing task to selected BSs, the traffic prediction process should begin about 100 ms ahead of the actual time when the sensing operation will be required. Traffic prediction on such short time scales can be done using ML models that learn the underlying behavior of traffic. We used the neural network to predict future PRB usage for the next 500 ms based on the previous PRB usage within the last 10 seconds, each sampled at 100 ms intervals. Therefore, the input to the neural network is set to the length of 100 and the output, 5.

To assess the performance of the neural network we have compared the absolute error in PRB prediction (i.e., the gap between the actual utilization and the predicted utilization). We tested the neural network using a measured network dataset for downlink traffic traces for a period of 2,250 seconds, where 100 PRBs are available. The actual and predicted PRB allocations are plotted in Fig. 6, where the insert shows that the predicted and actual values follow similar trends. The Cumulative Distribution Function (CDF) of error as absolute difference between actual and predicted PRB allocations demonstrates good prediction accuracy: for 80% of the test data instants the predicted PRB usage differed at most by 10 PRBs from the subsequent actual usage.

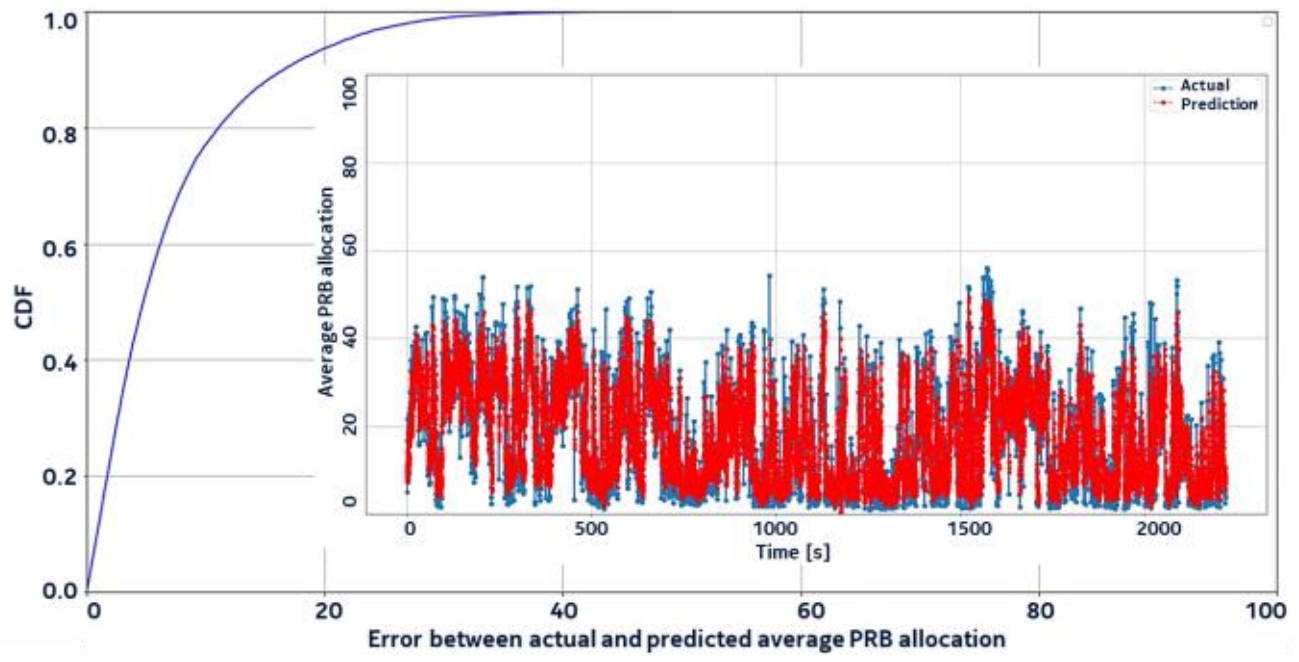

*Figure 6. CDF of error as absolute difference between actual and predicted PRB allocations over 2,250 seconds. Comparison between actual and predicted PRB utilization is shown in the insert.*

## VII. CONCLUSION

Viability of 6G in the upper mid-band requires an effective and efficient spectrum sharing mechanism that allows sufficiently high-power transmission to ensure similar coverage from the same site grid deployment as 5G. 6G technologies such as extreme MIMO facilitate much finer granularity sharing in the spatial domain thanks to the much narrower beamwidth from large antenna arrays. Scalable beamforming algorithms with adaptive power control can protect incumbents without substantially reducing cellular capacity. Additionally, AI-native support in 6G enables data-driven solutions for interference suppression and traffic prediction. Combined with effective spectrum sensing techniques, dynamic PRB blanking could be realized to share spectrum with satellite services.

One of the major challenges of data-driven optimization techniques is obtaining robust field involving the MN and the relevant incumbent systems. Standardizing frameworks for obtaining relevant datasets and appropriate sharing policy is important for the industry. Further research is also required on spectrum sharing based on spectrum sensing by cellular BSs. Research questions include how well the incumbent's potential use of spectrum can be predicted, how best to organize the spectrum sensing across the network, what kind of sensing accuracy can be achieved, and how fast the network can react to an incumbent starting to use their spectrum. Potential conflicts in requirements and resource allocation for different sharing mechanisms could also be explored. Further, regulatory authorities could encourage sharing of at least some limited, non-strategic information between satellite and terrestrial service providers to improve spectrum sensing accuracy and sharing efficiency. For military satellites that may include frequency hopping or frequency spreading, spectrum sensing will be more challenging, and some minimum information sharing would be needed to make sensing-based sharing a reality. Alternatively, a spectrum access server can inform the MN of the satellite's services impending use of frequencies in different geographical areas. The challenge here is to enable rapid dynamic spectrum sharing.

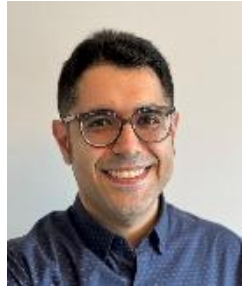

**Foad Sohrabi** (Member, IEEE) [foad.sohrabi@nokia-bell-labs.com] is a Member of Technical Staff in Nokia Bell Labs, Murray Hill, New Jersey.

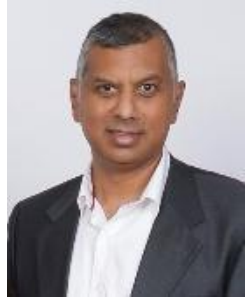

**Harish Viswanathan** (Fellow, IEEE) [harish.viswanathan@nokia-bell-labs.com] is the head of Radio Systems Research Lab, Nokia Bell Labs, Murray Hill, New Jersey.

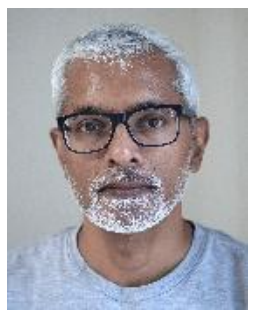

**Sivarama Venkatesan** (Member, IEEE) [venkat.venkatesan@nokia-bell-labs.com] is a Distinguished Member of Technical Staff in Nokia Bell Labs, Murray Hill, New Jersy.

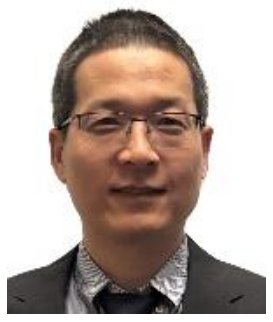

**Jinfeng Du** (Senior Member, IEEE) [jinfeng.du@nokia-bell-labs.com] is the department head of Radio Systems USA, Nokia Bell Labs, Murray Hill, New Jersey.

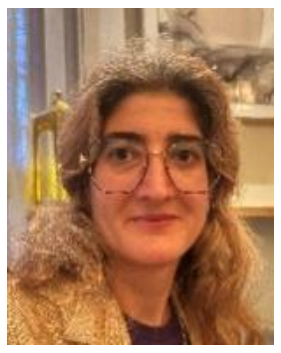

**Aliye Özge Kaya** (Senior Member, IEEE) [ozge.kaya@nokia-bell-labs.com] is a Distinguished Member of Technical Staff in Nokia Bell Labs, Murray Hill, New Jersey.

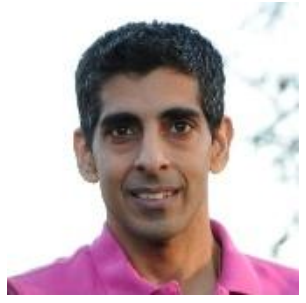

**Navin Hathiramani** (Member, IEEE) [navin.hathiramani@nokia.com] is a Principal Standardization Lead, Technology Standards, Dallas, Texas.

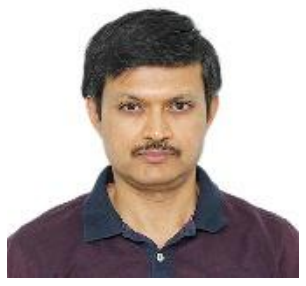

**Pavan Koteshwar Srinath** (Member, IEEE) [pavan.koteshwar_srinath@nokia-bell-labs.com] is a Research Engineer in Nokia Bell Labs, Messy, France.